\documentclass[final,journal]{IEEEtran}

\usepackage[acronym]{glossaries}
\newacronym{AB}{AB}{analog beamformer}
\newacronym{AP}{AP}{access point}
\newacronym{BF}{BF}{beamforming}
\newacronym{UE}{UE}{user equipment}
\newacronym{AWGN}{AWGN}{additive white gaussian noise}
\newacronym{MIMO}{MIMO}{multiple-input multiple-output}
\newacronym{MISO}{MISO}{multiple-input single-output}
\newacronym{RF}{RF}{radio frequency}
\newacronym{CL}{CL}{convolutional layer}
\newacronym{FDD}{FDD}{frequency division duplex}
\newacronym{TDD}{TDD}{time division duplex}
\newacronym{CSI}{CSI}{channel state information}
\newacronym{DNN}{DNN}{deep neural network}
\newacronym{DP}{DP}{digital precoder}
\newacronym{DL}{DL}{deep learning}
\newacronym{SVD}{SVD}{singular-value decomposition}
\newacronym{CNN}{CNN}{convolution neural network}
\newacronym{FDP}{FDP}{fully digital precoder}
\newacronym{SE}{SE}{spectral efficiency}
\newacronym{OFDM}{OFDM}{orthogonal frequency division multiplexing}
\newacronym{FL}{FL}{fully-connected layer}
\newacronym{HSHO}{HSHO}{Hybrid Structured Heuristic Optimization}
\newacronym{HBF}{HBF}{hybrid beamforming}
\newacronym{mm-Wave}{mm-Wave}{millimeter wave}
\newacronym{mMIMO}{mMIMO}{massive MIMO}
\newacronym{SINR}{SINR}{signal-to-interference-noise ratio}
\newacronym{SNR}{SNR}{signal-to-noise ratio}
\newacronym{SS}{SS}{synchronization signal}
\newacronym{SSB}{SSB}{synchronization signal burst}
\newacronym{RSSI}{RSSI}{received signal strength indicator}
\newacronym{PZF}{PZF}{phase zero forcing}
\newacronym{PSO}{PSO}{particle swarm optimization}
\newacronym{ZF}{ZF}{zero forcing}
\newacronym{O-FDP}{O-FDP}{optimal fully digital precoder}
\newacronym{JT}{JT}{joint transmission}
\newacronym{CU}{CU}{central unit}
\newacronym{CB}{CB}{conjugate beamforming}
\newacronym{NC}{NC}{network controller}
\newacronym{CoMP}{CoMP}{coordinated multi point}
\newacronym{CF-mMIMO}{CF-mMIMO}{cell-free massive MIMO}
\newacronym{CF-HBF}{CF-HBF}{cell-free hybrid beamforming}
\newacronym{CF-BF}{CF-BF}{cell-free beamforming}
\newif\ifDeepMIMOModel
\DeepMIMOModeltrue

\newif\ifSimpleNParamEq
\SimpleNParamEqtrue

\usepackage{lettrine}

\usepackage{ifpdf}

\ifCLASSINFOpdf
  \usepackage[pdftex]{graphicx}
  \graphicspath{{../pdf/}{../jpeg/}}
  \DeclareGraphicsExtensions{.pdf,.jpeg,.png}
\else
  \usepackage[dvips]{graphicx}
  \graphicspath{{../eps/}}
  \DeclareGraphicsExtensions{.eps}
\fi

\newcommand{\francois}[1]{\textcolor{black}{#1}}

\newcommand{\hamed}[1]{\textcolor{black}{#1}}

\usepackage{cite}
\usepackage{amssymb}
\usepackage{amsthm}
\usepackage{amssymb}
\usepackage{balance}
\usepackage{eqparbox}
\usepackage{multirow}
\usepackage[ruled,vlined]{algorithm2e}
\usepackage{adjustbox}
\usepackage{amsmath}

\usepackage{mathtools, nccmath}

\usepackage{booktabs, siunitx}
\usepackage[svgnames,table]{xcolor}

\usepackage{graphicx}
\usepackage[scaled]{DejaVuSansMono}
\usepackage[T1]{fontenc}

\usepackage{color}
\usepackage{relsize}
\usepackage{mathtools}

\newcommand{\mb}[1]{\mathbf{#1}}
\newcommand{\mr}[1]{\mathrm{#1}}

\DeclareMathOperator*{\minimize}{minimize}

\SetKwInput{KwInput}{Input}                
\SetKwInput{KwOutputr}{Output Regression}              
\SetKwInput{KwOutputc}{Output Classification}              

\newcommand{\bseq}{\begin{subequations}}
\newcommand{\eseq}{\end{subequations}}
\newcommand{\baln}{\begin{align}}
\newcommand{\ealn}{\end{align}}
\newcommand{\balnd}{\begin{aligned}}
\newcommand{\ealnd}{\end{aligned}}
\newcommand{\beq}{\begin{equation}}
\newcommand{\eeq}{\end{equation}}
\newcommand{\beqn}{\begin{eqnarray}}
\newcommand{\eeqn}{\end{eqnarray}}
\newcommand{\beqno}{\begin{eqnarray*}}
\newcommand{\eeqno}{\end{eqnarray*}}
\newcommand{\bma}{\begin{displaymath}}
\newcommand{\ema}{\end{displaymath}}
\newcommand{\bnu}{\begin{enumerate}}
\newcommand{\enu}{\end{enumerate}}
\newcommand{\bce}{\begin{center}}
\newcommand{\ece}{\end{center}}
\newcommand{\btb}{\begin{tabular}}
\newcommand{\etb}{\end{tabular}}
\newcommand{\ba}{\begin{array}}
\newcommand{\ea}{\end{array}}

\begin{document}

\title{\hamed{Decentralized Beamforming for Cell-Free Massive MIMO with Unsupervised Learning}}

\author{Hamed Hojatian,~\IEEEmembership{Member,~IEEE,}
        J\'{e}r\'{e}my Nadal,~\IEEEmembership{Member,~IEEE,}\\ Jean-Fran\c{c}ois Frigon,~\IEEEmembership{Senior Member,~IEEE,}
        and Fran\c{c}ois Leduc-Primeau,~\IEEEmembership{Member,~IEEE}
\vspace{-0.4cm}

}

\maketitle

\begin{abstract}
\Gls{CF-mMIMO} systems represent a promising approach to increase the spectral efficiency of wireless communication systems. However, near-optimal beamforming solutions require a large amount of signaling exchange between access points~(APs) and the network controller~(NC). In this letter, we propose two unsupervised deep neural networks~(DNN) architectures, fully and partially distributed, that can perform decentralized coordinated beamforming with zero or limited communication overhead between APs and NC, for both fully digital and hybrid precoding.
The proposed DNNs achieve near-optimal sum-rate while also reducing computational complexity by $10-24\times$ compared to conventional near-optimal solutions.
\end{abstract}


\IEEEpeerreviewmaketitle
\thispagestyle{empty}

\section{Introduction}\label{sec:intro}

\lettrine[lines=2]{C}{ell-Free} massive MIMO~(CF-mMIMO) networks
have the potential to significantly improve the efficiency of future wireless networks, as compared to cellular networks, by serving uniformly multiple users simultaneously using multi-antenna \glspl{AP} connected to a central \gls{NC}~\cite{marzetta,7827017, 8417645}. 
Similar to standard \gls{mMIMO} systems, \gls{CF-mMIMO} requires designing suitable precoders for data transmission, with the added challenge that information exchange between \glspl{AP} and \gls{NC} should be minimized~\cite{8845768}. Existing techniques tend to exhibit a trade-off in that regard.
For instance, in the context of \gls{FDP}, the simple and scalable \gls{CB} method can be implemented locally by each \gls{AP} and achieves acceptable performance without information exchange~\cite{7917284}. On the other hand, the \gls{ZF} method achieves much better performance, but the precoders are computed centrally in the \gls{NC} at the expense of fronthaul overhead~\cite{7917284}. 

\Gls{HBF} is a well-known approach to reduce energy consumption by decreasing the number of \gls{RF} chains in the transmitter without reducing the number of antennas~\cite{hbf_survey}.
However, designing \gls{HBF} precoders that achieve near-optimal performance usually has a high computational cost.
Several works have investigated the use of \gls{DL} to design the \gls{HBF} for single-cell communication~\cite{hojatian2020unsupervised, HBF}, but extending these solutions to \gls{CF-mMIMO} imposes a large signaling overhead between \glspl{AP} and \gls{NC} to exchange the beamforming information.
In~\cite{alkh2}, the authors proposed a supervised deep learning-based beamforming design for coordinated beamforming. However, they consider that the beamforming vectors are designed centrally in the \gls{NC}, and only consider a simplistic analog-only beamforming scenario for a single user with one \gls{RF} chain per base station.

Moreover, all mentioned studies either assume that the \gls{CSI} is known or the system works in \gls{TDD} where perfect channel reciprocity exists. However, in practice, the channel reciprocity may not be accurate because of the calibration error in the RF chains, hardware impairment issues, or time-varying channel~\cite{7166317}. 

In this letter, we consider a \gls{FDD} \gls{CF-mMIMO} system with multiple \glspl{AP}, each equipped with \gls{HBF}, cooperatively serving multiple users simultaneously.
We propose distributed unsupervised \gls{DL}-based solutions to perform decentralized HBF cooperatively and we show that appropriate training of the \glspl{DNN} allows eliminating all fronthaul signaling overhead during the online phase. 
Through simulations based on the \emph{deepMIMO} ray-tracing model~\cite{deepmimo}, we show that the proposed solution can achieve near-optimal sum-rate performance with reduced complexity compared to existing approaches.
We also provide an example of the trade-off between overall computational complexity and signaling overhead by designing an alternative architecture for which complexity is further reduced at the cost of increased fronthaul signaling.
In addition, we show that the proposed schemes can also be used to reduce the computational complexity and signaling overhead in a coordinated \gls{FDP} system.
All solutions are based on our previously proposed unsupervised learning method \cite{hojatian2020unsupervised} that avoids the need to provide examples of known optimal solutions. 

The remainder of this letter is organized as follows. 
Section~\ref{secII:system-model} describes the system model. 
Section~\ref{secIII:baseline} presents the non-DL \gls{CF-HBF} method used as a baseline.
Section~\ref{sec:IV} presents the proposed DNN architectures and algorithms. 
Numerical results are provided in Section~\ref{sec:simulation}, followed by a conclusion in Section~\ref{sec:Conclusion}.

\section{System Model}\label{secII:system-model}


We consider a \gls{CF-mMIMO} network, where $M$ \glspl{AP} each equipped with $N_{\sf{T}}$ antennas
communicate with a \gls{NC} through a fronthaul connection, while serving simultaneously $N_{\sf{U}}$ single antenna users. Each \gls{AP} is assumed to have $N_{\sf{RF}} << N_{\sf{T}}$ RF chains. The signal received by each user is
\begin{equation}\label{eq:sys_model}
    y_u = \underbrace{\sum _{\forall m} \mb{h}_{u,m}^{\rm{H}} \mb{A}_{m} \mb{w}_{u,m} x_{u}}_{\text{Desired signal}} + \underbrace{\sum _{\forall m} \mb{h}_{u,m}^{\rm{H}} \mb{A}_{m} \sum_{j \neq u} \mb{w}_{j,m}x_{j}}_{\text{Interference}} + \eta_{u},
\end{equation}
where $x_{u}$ is transmit symbol for user index $u$, $\mb{h}_{u,m}$ is the channel vector between the $u^{\text{th}}$ user and $m^{\text{th}}$ \gls{AP},  $\mb{A}_{m} \in  \{1,-1,i,-i\}^{N_{\sf{T}} \times N_{\sf{RF}}}$ is the \gls{AB} selected from the $m^{\text{th}}$ codebook ($\mathcal{A}_{m}$), $\mb{w}_{u,m} \in \mathbb{C}^{N_{\sf{U}} \times 1}$ is the \gls{DP} for the $u^{\text{th}}$ user, and $\eta_{u}$ is the zero-mean Gaussian noise with variance $\sigma^2$. The \gls{SINR} for the $u^{\text{th}}$ user is given by
\begin{eqnarray}\label{eq:SINR}
\mr{SINR}_{u}(\mb{A}, \mb{W}) = \frac{ \big|\sum _{\forall m} \mb{h}_{u,m}^{\rm{H}} \mb{A}_{m} \mb{w}_{u,m} \big|^2}{\big| \sum _{\forall m} \mb{h}_{u,m}^{\rm{H}} \mb{A}_{m} \sum_{j \neq u} \mb{w}_{j,m} \big|^2 + \sigma^2} \,,
\end{eqnarray}
where the global \gls{AB} is defined as the block diagonal matrix $\mb{A} = \text{diag}(\mb{A}_1, ..., \mb{A}_M)$ because independent \glspl{AP} are deployed, and the global \gls{DP} is defined as $\mb{W} = [\mb{W}_1, \cdots, \mb{W}_M]^{\rm T}$. The sum-rate of the system is therefore
\beq \label{eq:sumRate_HBF}
R_\text{HBF}(\mb{A}, \mb{W}) = \sum_{\forall u} \log_2(1 + \mr{SINR}_{u}(\mb{A}, \mb{W})) \, .
\eeq
We focus on \gls{CF-HBF} design to maximize the sum-rate corresponding to the following optimization problem:
\begin{subequations}  \label{SEE-max-prb}
\begin{eqnarray} 
&\underset{\left\lbrace  \mb{A} ,\mb{W}\right\rbrace}{\max} & R_{\text{HBF}}(\mb{A}, \mb{W}) = \sum_{\forall u} \log_2(1 + \mr{SINR}_{u}(\mb{A}, \mb{W})) \\
& \text{s.t.} &  \sum_{\forall m} \mb{w}^{\rm{H}} _{u, m} \mb{A}^{\rm{H}}_{m}  \mb{A}_{m} \mb{w}_{u, m} \leq P_{\sf{max}}, \label{cnt2} \\
&& \mb{A}_{m} \in \mathcal{A}_{m} \label{cnt3},
\end{eqnarray}
\end{subequations}
where $P_{\sf{max}}$ stands for the total maximum transmission power in the \gls{CF-mMIMO} network. In this paper, without loss of generality, we consider $P_{\sf{max}}=~1$.

\label{sec:baseline_HBF}
\subsection{Beam Training}
Beam training is required for initial access and to obtain \gls{CSI} between the \glspl{AP} and the users. CSI acquisition is a challenging task for \gls{mMIMO} system especially in \gls{FDD} communication. 
We therefore suggested in~\cite{hojatian2020rssi} a beam training method that relies on simpler \gls{RSSI} feedback instead of explicit \gls{CSI}.  Our proposed beam training for \gls{CF-mMIMO} follows a similar approach as in the single-cell case described in~\cite{hojatian2020unsupervised}. However, in the \gls{CF-mMIMO} beam training, each \glspl{AP} takes a turn sending its \gls{SS}, and each user measures the \glspl{RSSI} from each \glspl{AP}.

In the first step each \gls{AP} transmits $K$ \gls{SSB} sequentially, where each burst $k \in [1,K]$ uses a different analog-only beamforming $\mb{a}_{\sf{SS},m}^{(k)} \in  \{1,-1,i,-i\}^{N_{\sf{T}} \times 1}$. The \glspl{SSB} are designed for each \glspl{AP} individually for initial access using the method proposed in~\cite{hojatian2020unsupervised}. The synchronization signal $\mb{a}_{\sf{SS},m}^{(k)}$ sent by the $m^{\text{th}}$ \gls{AP} in the downlink channel is received by all users. Therefore, the received signal $r_{u,m}^{(k)}$ at the $u^{\text{th}}$ user for the $k^{\text{th}}$ burst from the $m^{\text{th}}$ \gls{AP} is $r_{u,m}^{(k)} = \mb{h}^{\rm{H}}_{u,m} \mb{a}_{\sf{SS},m}^{(k)} + \eta^{(k)}_u.$

In the next step, the \gls{RSSI} values $\alpha_{u,m}^{(k)}$ are measured by the $u^{\text{th}}$ user for the $k^{\text{th}}$ \gls{SS} burst as
$\alpha_{u,m}^{(k)} = \left|  r_{u,m}^{(k)}  \right| ^2  + \sigma^2.$
Then, each user sends a set of measured \glspl{RSSI} ($\alpha_{u,m} = [\alpha_{u,m}^{(1)}, ..., \alpha_{u,m}^{(K)}]$) through a dedicated error-free feedback channel to the corresponding \gls{AP}. Therefore each user sends back $M \times K$ \gls{RSSI} values, and the \glspl{RSSI} received by $m^{\text{th}}$ \gls{AP} is $\alpha_m = [\alpha_{1,m}, ..., \alpha_{N_{\sf{u}},m}]^{\text{T}}$ with dimension $N_{\sf{U}} \times K$.

%

%

\subsection{Codebook Design}
\label{subsec:core_DNN_datasets}

The number of possible \gls{AB} phase combinations as codeword grows exponentially with the number of antennas and RF chains. However, for a given channel environment, only a subset of these combinations are useful, and the search space can be highly reduced. A 3-step codebook design is proposed in~\cite{hojatian2020unsupervised}. In this letter, we deployed the PE-AltMin algorithm proposed in~\cite{7397861} in second step to find the optimal \gls{AB} solutions. 
Then, the codebook size is iteratively reduced by discarding the less-used \gls{AB} solutions in the codebook. In this paper, we design the codebook $\mathcal{A}_m$ for each \gls{AP} individually using a similar approach. Thus, the size of the codebook for each \gls{AP} would be different because it depends on the \gls{AP}'s location and on its channel environment.


\section{Baseline Cell-Free Hybrid Beamforming with Perfect CSI}\label{secIII:baseline}

According to \eqref{eq:SINR}, a fully connected \gls{CF-HBF} can be seen as a single  \gls{mMIMO} cell equipped with $M \times N_{\sf{T}}$ antenna and $M \times N_{\sf{RF}}$ RF chains. 
Therefore, the general approach is to first jointly design the \gls{FDP} for all \gls{AP}. Then, the \gls{AB} and \gls{DP} are designed independently for each of the $M$ APs using

\beq 
\underset{\mb{A}_m}{\minimize}  \big\|\mb{U}_m - \mb{A}_m \mb{W}_m\big\|^2 \hspace{8mm}
\text{s.t. \eqref{cnt2}, \eqref{cnt3}}. \label{MMSE_prb}
\eeq
where $\mb{U} = [\mb{U}_1, ..., \mb{U}_M]$ is the global \gls{FDP} matrix, $\mb{U}_m = [\mb{u}_{1,m}, ..., \mb{u}_{N_{\sf{U}},m}]^{\text{T}}$ is the fully digital precoder for \gls{AP} index $m$ and $\mb{u}_{u,m} \in \mathbb{C}^{N_{\sf{T}} \times 1} $ is the fully digital precoder vector in the $m^{\text{th}}$ \gls{AP} for user index $u$. To obtain the \gls{FDP} solution we define the optimization problem over all \glspl{AP} as,

\begin{subequations} \label{eq:FDP_opti_problem}
\begin{eqnarray} 
&\underset{\left\lbrace \mb{U} \right\rbrace}{\max} &
\sum_{\forall u}  R_\text{FDP}(\mb{U})  \\
& \text{ s. t.} & \sum_{\forall u} \mb{U}^{\rm{H}} \mb{U} \leq P_{\sf{max}}, \label{cnt_op2}
\end{eqnarray}
\end{subequations}
where $R_\text{FDP}(\mb{U}) = \sum_{\forall u} \log_2(1 + \mr{SINR}_{u}(\mb{U})) \,$ and
\begin{eqnarray}\label{eq:FDP-SINR}
\mr{SINR}_{u}(\mb{U}) = \frac{ \big|\sum _{\forall m} \mb{h}_{u,m}^{\rm{H}} \mb{u}_{u,m} \big|^2}{\big| \sum _{\forall m} \mb{h}_{u,m}^{\rm{H}} \sum_{j \neq u} \mb{u}_{j,m} \big|^2 + \sigma^2} \,.
\end{eqnarray}

The method we employed to find the \gls{O-FDP} is based on~\cite{BBO2014}, where it is demonstrated that the \gls{O-FDP} vector $\mb{u}_{u,m}$ of \gls{FDP} matrix $\mb{U}_m$ for user  $u$  has the following analytical structure:
%
%
\begin{align}
    \mb{u}_{u,m} = \sqrt{p_u} \frac{\Bigl( I_{N_{\sf{U}}} + \frac{1}{\sigma^2}\sum_{i=1}^{N_{\sf{U}}-1}  \mb{h}_{i,m}\lambda_i\mb{h}^{\rm{H}}_{i,m} \Bigr)^{-1} \mb{h}_{u,m}}{ \left\Vert \Bigl( I_{N_{\sf{U}}} + \frac{1}{\sigma^2}\sum_{i=1}^{N_{\sf{U}}-1}  \mb{h}_{i,m}\lambda_i\mb{h}^{\rm{H}}_{i,m} \Bigr)^{-1} \mb{h}_{u,m} \right\Vert}, \label{FDP_ana}
\end{align}
where $I_{N_{\sf{U}}}$ corresponds to the $N_{\sf{U}} \times N_{\sf{U}}$ identity matrix, and $p_u$ and $\lambda_u$ are the unknown real-valued coefficients to be optimized, respectively corresponding to the beamforming power and Lagrange multiplier for user $u$. Once $p_u$ and $\lambda_u$ have been found, \eqref{FDP_ana} can be substituted into \eqref{MMSE_prb}, and this optimization problem can be solved using the PE-AltMin solution proposed in~\cite{7397861}. We consider this solution~(PE-AltMin) as a near-optimal baseline solution to evaluate the proposed \gls{DNN}-based architectures. However, this near-optimal method is difficult to implement in real-time systems due to its heavy computational complexity. 
Furthermore, it depends on having the full \gls{CSI} and it is a centralized method, where the \gls{HBF} vectors are computed in the \gls{NC} and then sent to each \gls{AP}, thus requiring high capacity fronthaul links.

\section{\hamed{Distributed DNNs for Cell-Free Beamforming}} \label{sec:IV}

\hamed{We propose two possible architectures for DNN-based cell-free beamforming (HBF or FDP), each achieving a different trade-off between computational complexity and signaling overhead. 
In the first architecture, fully decentralized beamforming, the trained \gls{DNN} is fully distributed and the \gls{NC} does not participate in beamforming design. 
In the second architecture, partially decentralized beamforming, only last two layers of the \gls{DNN} is distributed at each \gls{AP}, and the \gls{NC} remains involved in the online phase for beamforming design.}

\subsection{\hamed{Fully Decentralized Beamforming}} \label{sub-FullyDC-Net}
\begin{figure}[t!]
    \centering
    \includegraphics[width=\columnwidth]{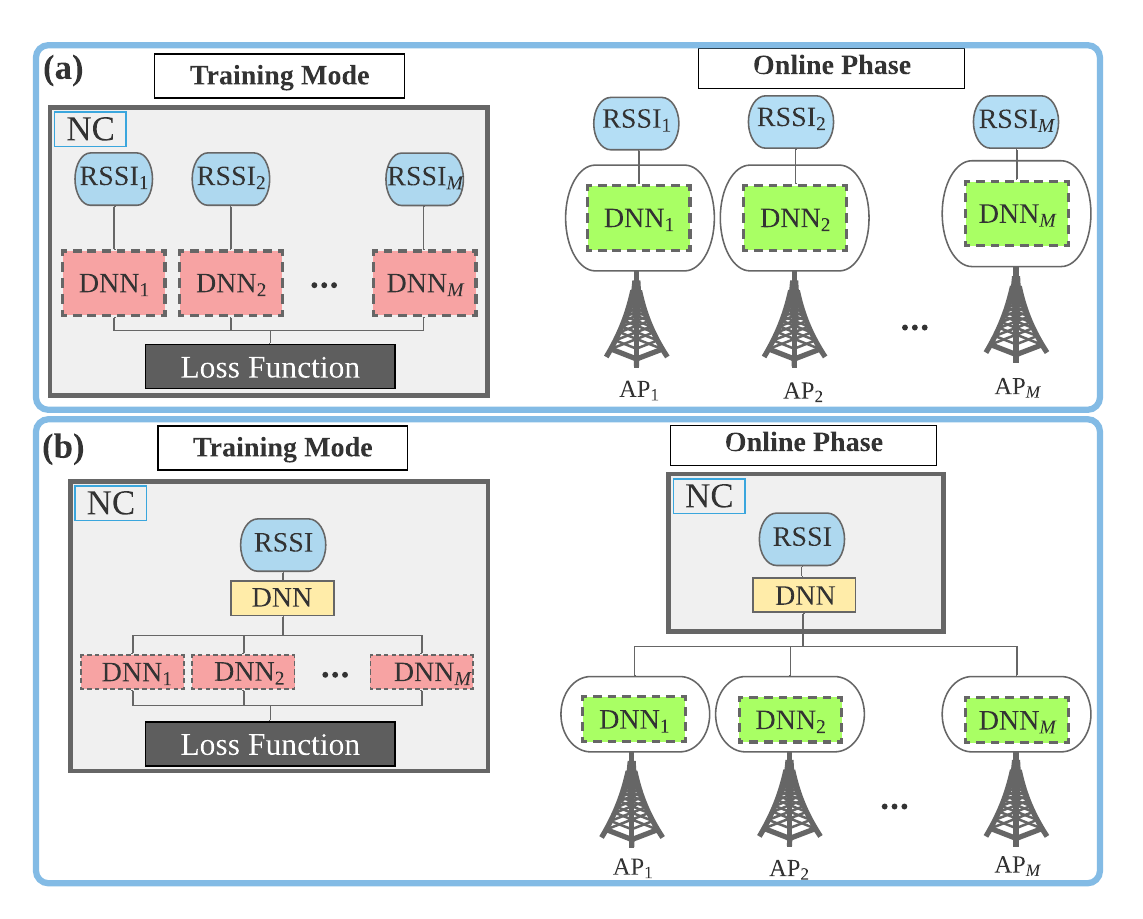}
    \caption{a) fully distributed DNN ``FullDeC-HBF'' or ``FullDeC-FDP'', b) partially distributed DNN ``PartDeC-HBF'' or ``PartDeC-FDP''.}
    \label{fig:arch_CNN_FD}
\end{figure}

The proposed fully distributed architecture,  called ``FullDeC-HBF'' for HBF \hamed{or ``FullDeC-FDP'' for FDP}, is shown in Fig~\ref{fig:arch_CNN_FD}-(a). The main idea is to completely transfer the signaling exchange of the beamforming from the online mode to the training phase. To do so, the architecture is composed of $M$ parallel local-\glspl{DNN}, each taking as input only the RSSI associated with one AP. These networks are trained jointly, but during the online mode, each AP uses only its trained local-DNN, and designs its beamforming vector locally, which eliminates the fronthaul signaling overhead. 

\hamed{In \emph{FullDeC-HBF},} a multi-tasking \gls{DNN} is considered, which jointly performs the regression and classification task to respectively design the \gls{DP} and the \gls{AB}. Each local-\gls{DNN} consists of $2$ \glspl{CL} with $32$ channels, followed by $2$ \glspl{FL} with $512$ neurons connected to the output layer. 
Since we use real-valued \glspl{DNN}, 
the output layer for the regression task has size $2 \times N_{\sf{RF}} \times N_{\sf{U}}$ for each local-DNN.
All non-output layers use the ``LeakyReLU’' activation function and for the output layer of the classifier, the ``Softmax'' activation function is used to assign a probability to each codewords in the codebook. \hamed{Hence, we define $\mb{p}_m = [p_{1,m}, ..., p_{l,m}, ..., p_{L_m,m}]$ as the output of each classifier, where $p_{l,m}$ corresponds to the probability of the $l^{\text{th}}$ codeword in the codebook of the $m^{\text{th}}$ \gls{AP} and $L_m = |\mathcal{A}_m|$ is the size of $m^{\text{th}}$ codebook. The size of the classifier in each local-DNN corresponds to the length of the local codebook. 
For the FDP case, the architecture of \emph{FullDeC-FDP} is the same as \emph{FullDeC-HBF}, except for the output layer which only consists of a regression task since there is no AB.}

Since we design $M$ parallel local-\glspl{DNN}, the complexity linearly scales by increasing the number of \glspl{AP}. To address this concern, we propose another architecture in the following based on the auto-encoder concept, enabling lower computational complexity than fully decentralized beamforming.

\subsection{\hamed{Partially Decentralized Beamforming}}

The second architecture, called ``PartDeC-HBF'' for HBF \hamed{or ``PartDeC-FDP'' for FDP, is shown in Fig~\ref{fig:arch_CNN_FD}-(b).} Here, we designed the \gls{DNN} partially distributed with a combination of shared and unshared layers in the training phase. The first idea behind this architecture is to use some shared layers to reduce the total computational complexity both in the training phase and the online phase. \hamed{To do so, we used $2$ shared \glspl{CL} with $32$ channels, one \gls{FL} with $1024$ neurons followed by $M$ parallel groups of unshared layers. Each group consists of $2$ \glspl{FL} with $200$ and $1024$ neurons.} 
\hamed{The second idea is to exploit the fact that the combination of the last shared layer with a first unshared layer acts like an auto-encoder, enabling to reduce the signaling overhead.} 
The activation functions are the same as in the fully decentralized case.
\hamed{For the FDP case, \emph{PartDeC-FDP} is derived from \emph{PartDeC-HBF} by keeping only the regression task in the output layer.}

\begin{table*}[t!]
    \centering
    \caption{\hamed{Comparison of Beamforming Type, Signaling Exchange, and Complexity ($N_{\sf{T}} = 64$, $N_{\sf{RF}} = 8$, $N_{\sf{U}} = 4$, $K=16$, $M=4$) }}
    \resizebox{0.9\textwidth}{!}{
    \begin{tabular}{l|ccccccc}
        \toprule
        \multicolumn{1}{c}{}& 
        \multicolumn{1}{c}{Beamforming}&
        \multicolumn{1}{c}{\# RF chains}&
        \multicolumn{2}{c}{Signaling Exchange} &
        \multicolumn{1}{c}{\# multiplications} & \multicolumn{1}{c}{Sum-Rate} &
        \multicolumn{1}{c}{Architecture} \\
        \multicolumn{1}{c}{Technique} & Type & {(per AP)} & APs $\to$ NC & NC $\to$ APs &($\times 10^6)$ & (bit/s/Hz) & Type   \\
        \cmidrule(lr){1-1} \cmidrule(lr){2-2} \cmidrule(lr){3-3} \cmidrule(lr){4-5} \cmidrule(lr){6-6} \cmidrule(lr){7-7} \cmidrule(lr){8-8}
        
        ZF~\cite{7917284}  & FDP (Perfect CSI) & $ N_{\sf{T}}$  & $2MN_{\sf{T}}N_{\sf{U}}$ & $2 M  N_{\sf{T}} N_{\sf{U}}$ &   24.4 & 25.4 (100\%) & Centralized\\
        \hamed{\textbf{FullDeC-FDP}} & FDP (RSSI-based) & $N_{\sf{T}}$  & 0 & 0 & 2.9 & \textbf{23.3} (92\%)& Decentralized \\
        \hamed{\textbf{PartDeC-FDP}} & FDP (RSSI-based) & $N_{\sf{T}}$  & $KM N_{\sf{U}}$ & $200 M$ & 1.3 & \textbf{23.2} (92\%) & Centralized\\
        CB~\cite{7917284} & FDP (Perfect CSI) & $N_{\sf{T}}$  & 0 & 0 & 0 & 13.1 (52\%) &Decentralized\\
        \cmidrule(lr){1-1} \cmidrule(lr){2-8}
        PE-AltMin~\cite{7397861} + O-FDP & HBF (Perfect CSI) & $N_{\sf{RF}}$  & $2MN_{\sf{T}}N_{\sf{U}}$ & $2 M N_{\sf{RF}} ( N_{\sf{T}} + N_{\sf{U}})$ &  $  369.2$ & 20.2 (100\%) & Centralized\\
        PE-AltMin~\cite{7397861} +  ZF\cite{7917284} & HBF (Perfect CSI) & $N_{\sf{RF}}$  & $2MN_{\sf{T}}N_{\sf{U}}$ & $2 M N_{\sf{RF}} ( N_{\sf{T}} + N_{\sf{U}})$ &  $26.2$ & 19.7 (97\%) & Centralized\\
        \textbf{FullDeC-HBF} & HBF (RSSI-based) & $N_{\sf{RF}}$  & 0 & 0 & 2.7 & \textbf{19.5} (96\%)& Decentralized \\
        \textbf{PartDeC-HBF} & HBF (RSSI-based) & $N_{\sf{RF}}$  & $KM N_{\sf{U}}$ & $200 M$ & \hamed{1.1} & \hamed{\textbf{19.4} (96\%)} & Centralized\\
        \bottomrule
    \end{tabular}}
    \label{tbl:signaling}
\end{table*}

\subsection{Training Mode}
As shown in Fig\ref{fig:arch_CNN_FD}-(a),  all local-\glspl{DNN} in \emph{FullDeC-HBF} or \hamed{\emph{FullDeC-FDP} are} trained jointly, for instance inside the \gls{NC},
and all of them are fed with quantized \glspl{RSSI} obtained from users, as described in Section~\ref{secII:system-model}. 
Since we aim to train the \gls{DNN} with unsupervised learning, we propose the following loss function to train the \glspl{DNN} for \gls{HBF}:
\beq \label{loss_HBF_local}
\mathcal{L}_{\text{HBF}} = - \sum_{l_1 = 1}^{L_1}  ... \sum_{l_M=1}^{L_M} \Bigg(  R_\text{HBF}(\Bar{\mb{A}}_{l_1, ..., l_M},\Bar{\mb{W}}) \prod_{m=1}^{M} p_{l_m,m}  \Bigg),
\eeq
where $\bar{\mb{W}}$ is the \gls{DP} output of the \gls{DNN},  
\hamed{$\Bar{\mb{A}}_{l_1,...,l_M} = \text{diag}(\Bar{\mb{A}}_{l_1},...,\Bar{\mb{A}}_{l_M})$, and $\Bar{\mb{A}}_{l_m}$ is the AB corresponding to the codeword index $l_m$. Thus, the loss function is defined in terms of the expected sum rate of the system, given the probabilities assigned to each codeword combination for all local DNNs. Likewise, we define $\mathcal{L}_{\text{FDP}} = - R_\text{FDP}(\mb{\Bar{U}})$ for the FDP where $\mb{\Bar{U}}$ is the output of the DNN in \emph{FullDeC-FDP}.}

Moreover, batch normalization and dropout are used during training. Finally, to satisfy $\sum_{\forall m} ||\Bar{\mb{A}}_{l_m} \Bar{\mb{w}}_m||^2 = 1$, we further normalize $\Bar{\mb{w}}_{m}$ using the approach proposed in~\cite{7917284} for power allocation.  The training phase for \emph{PartDeC-HBF} \hamed{or \emph{PartDeC-FDP}} shown in Fig~\ref{fig:arch_CNN_FD}-(b) follows the same procedures.

\subsection{Online Mode}
In the evaluation phase, each AP is assumed to have a local copy of a portion of the DNN (green boxes in Fig.~\ref{fig:arch_CNN_FD}).
In the case of the fully distributed systems (\emph{FullDeC-HBF} or \emph{FullDeC-FDP}), the local DNNs are fully independent, and each \gls{AP} is able to directly design the precoding as soon as it receives its quantized RSSI feedback.
In the case of the partially distributed systems (\emph{PartDeC-HBF} or \emph{PartDeC-FDP}), the local DNNs receive inputs from the shared DNN layers (yellow box) in the NC.
Consequently, the quantized RSSI input is first processed by the NC, and then the real-valued outputs of the shared DNN are sent from the NC to the APs, which then evaluate the last layers to output the precoding. Note that, although the \gls{NC} is engaged in the online mode, the signaling overhead is nonetheless reduced compared to a conventional method since the last layer in the NC and the first layer in each AP are sized to form an auto-encoder.

\section{Simulation Results} \label{sec:simulation}
In this section, the performance of the \hamed{four} proposed \hamed{architectures (two for HBF and two for FDP)}, implemented using the \textsc{PyTorch} \gls{DL} framework, is evaluated numerically\footnote{Code is available at https://github.com/HamedHojatian/CF-mMIMO-HBF.}. The deepMIMO channel model~\cite{deepmimo} is employed to generate the dataset, with parameters $\texttt{active\_BS}= \{4:10 \}$, $\texttt{active\_user\_first} = 1100$ and $\texttt{active\_user\_last} = 2200$. 
There are $M$ APs ($M \in \{2,4,8\}$), each equipped with $N_{\sf{T}} = 64$ antennas and $N_{\sf{RF}} = 8$ RF chains with 2-bit phase shifters serving $N_{\sf{U}} = 4$ users located randomly. The SSB has size $K=16$, and the RSSI feedback values are quantized on $8$ bits.\footnote{See \cite{hojatian2020unsupervised} for an evaluation of the impact of RSSI quantization on performance in the single AP case.}
The size of the DNN dataset is set to $10^6$ samples, with $85$\% of the samples used for the training set and the remaining ones used to evaluate the performance as test set. The mini-batch size, learning rate and weight decay are set to $1000$, $0.001$, and $10^{-6}$, respectively.


Table~\ref{tbl:signaling} compares the amount of signaling exchange, the computational complexity, and the sum-rate performance of the proposed methods with existing approaches for $\sigma^2 = -130$~dBW. \hamed{The top and bottom rows of Table~\ref{tbl:signaling} respectively present results for \gls{FDP} and \gls{HBF} techniques}. The amount of signaling exchange between the \glspl{AP} and NC is found by counting the number of transferred real matrix coefficients. For the computational complexity, we consider the number of real multiplications~(RM) for each matrix multiplication and inversion involved in the algorithms. We assume that one complex multiplication~(CM) corresponds to $4$ RMs. General expressions for the number of RMs required by O-FDP and by each DNN layer can be found in~\cite{hojatian2020unsupervised}. 
\hamed{In order to fairly compare the complexity of the fully and partially decentralized alternatives, the DNNs are sized such that the ``Full'' and ``Part'' DNNs achieve approximately the same sum rate.}

As a near-optimal baseline for HBF, we adapted \emph{PE-AltMin}~\cite{7397861} for the case of 2-bit phase shifters. 
For \gls{FDP}, we compare the proposed solutions with \gls{ZF} and \gls{CB}. Since the \emph{PE-AltMin} method is based on knowing the \gls{FDP} matrix, Table~\ref{tbl:signaling} considers both a high complexity near-optimal approach (O-FDP) and a low complexity approach (ZF) for obtaining the FDP.  The average number of iterations for \emph{PE-AltMin} to converge is $\ell=18$. Therefore, considering that the singular-value decomposition of an $m \times n$ matrix requires $4m^2 n + 22n^3$ RMs~\cite{Golub1996}, the number of RMs for \emph{PE-AltMin} can be expressed as $\ell M(8N_{\sf{RF}}N_{\sf{U}}(N_{\sf{T}} + N_{\sf{U}}) + 22 N_{\sf{RF}}^3)$. 

When compared to the \emph{PE-AltMin} + \gls{ZF} technique, \emph{PartDeC-HBF} has a slight sum-rate loss of \hamed{$1$\%}, but requires \hamed{$80$\%} less signaling exchange~(uplink + downlink), and is \hamed{$24\times$} less complex. Moreover, perfect CSI is used for all reference approaches, whereas the proposed DNNs only rely on RSSI measurements as described in Section~\ref{secII:system-model}. On the other hand, \emph{FullDeC-HBF} \hamed{also} has a sum-rate loss of \hamed{$1$\%}, but requires no signaling exchange while being almost $10\times$ less complex. In comparison to the \emph{PE-AltMin} + \gls{O-FDP}, \emph{FullDeC-HBF} and \emph{PartDeC-HBF} are respectively $136\times$ and $335\times$ less complex \hamed{at the cost of a $4$\% sum-rate reduction compared to \emph{PE-AltMin} + \gls{O-FDP}}. Therefore, both proposed DNNs provide near-optimal \gls{HBF} solutions with significantly less computational complexity and signaling exchange than traditional methods.


As expected, in FDP solution, \gls{ZF} provides the best FDP sum-rate in high SNR regime. The complexity of \gls{ZF} is given by $4M^2N_{\sf{T}}^2(2N_{\sf{U}} + MN_{\sf{T}}/3)$. \hamed{However, \emph{FullDeC-FDP} and \emph{PartDeC-FDP} have almost $8\times$ and $18\times$ lower computational complexity than \gls{ZF}, respectively, at the cost of $8$\% sum-rate loss.
For \emph{PartDeC-FDP}, the signaling exchange between \glspl{AP} and the \gls{NC} is reduced by $60$\% when compared to the \gls{ZF} solution, while there is no signaling exchange for \emph{FullDeC-FDP}. \gls{CB} is the less complex of all techniques and requires no signaling overhead. However, both proposed DNN FDP solutions outperform \gls{CB} with $77$\% higher sum-rate. It is worth noting that the FDP techniques require one RF chain per antenna~($64$ RF chains). Thus, they are not energy efficient compared to HBF techniques which only require $8$ RF chains} coupled with 2-bit phase shifters.

\begin{figure}
    \centering
    \includegraphics[width=1\columnwidth]{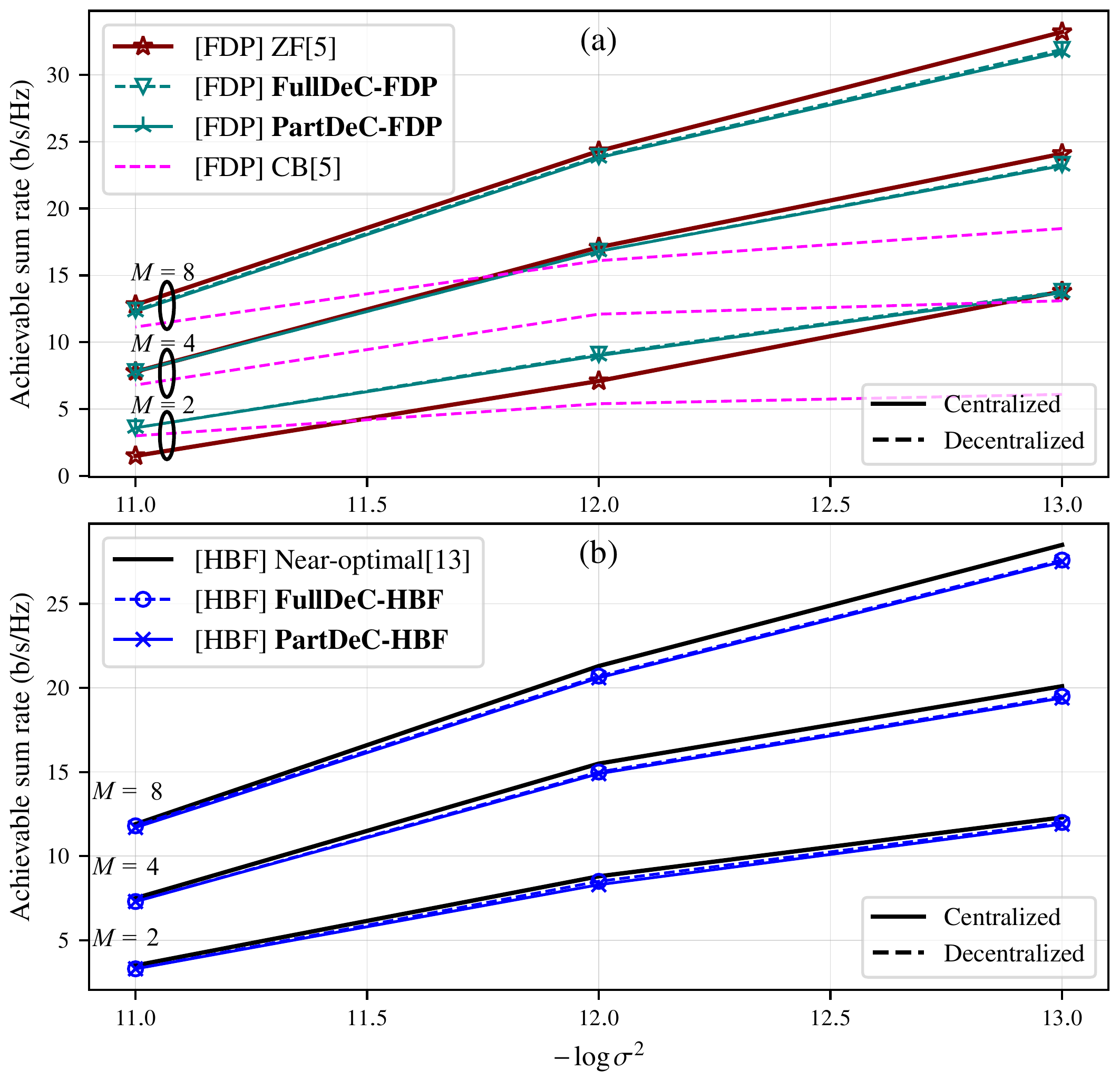}
    \caption{\hamed{Achievable sum-rate~(b/s/Hz) performance (a) FDP, (b) HBF.}}
    \label{fig:result_compnocom}
\end{figure}
\begin{figure}
    \centering
    \includegraphics[width=\columnwidth]{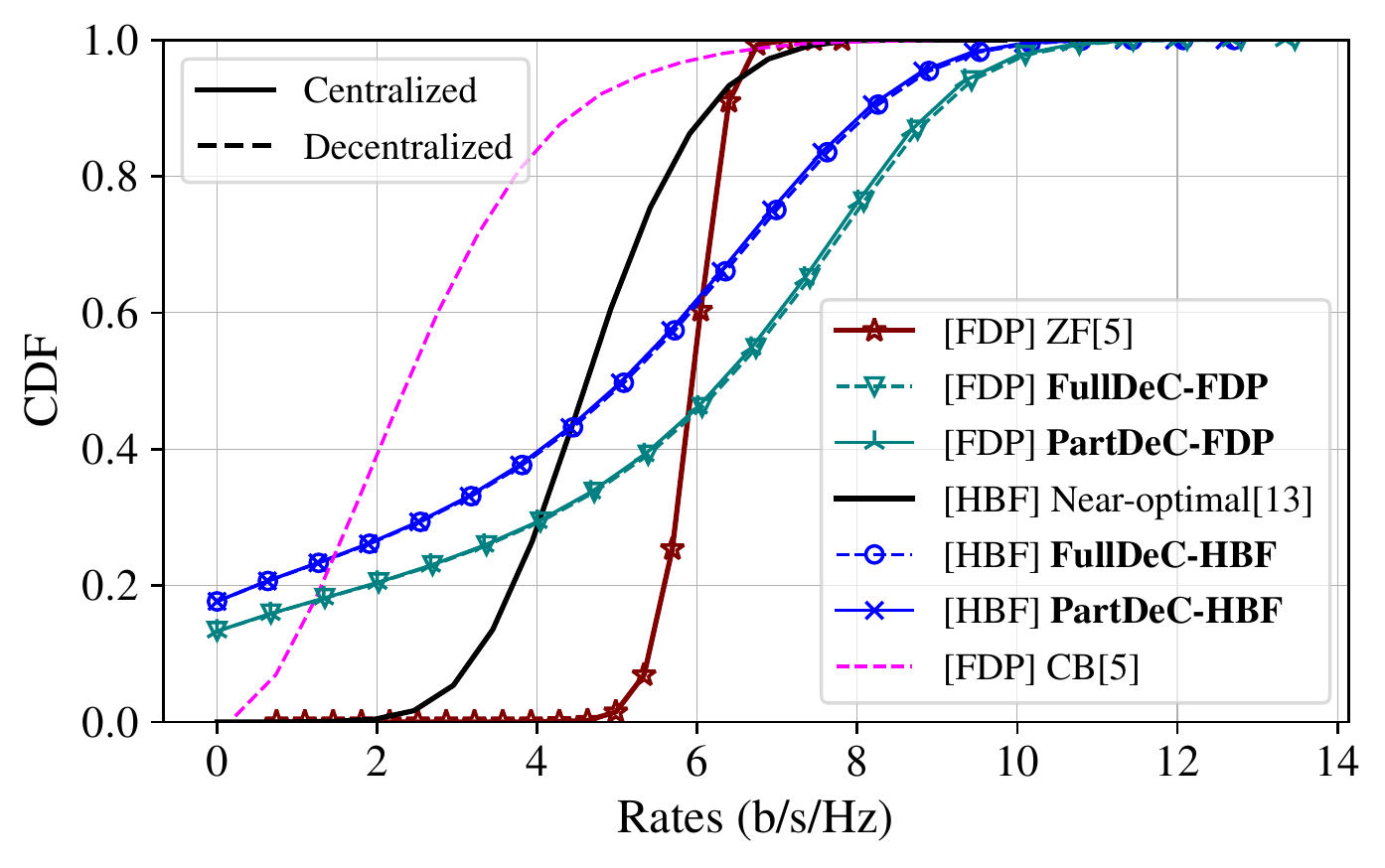}
    \caption{ \hamed{CDFs of the  per-user rates~($\sigma^2 = -130$, $M = 4$, $N_{\sf{T}} = 64$).}}
    \label{fig:result_CF_CDF}
\end{figure}

In Fig.~\ref{fig:result_compnocom} (a) and (b), we evaluated the achievable sum-rate of the proposed architectures for FDP and HBF, respectively, where in Fig.~\ref{fig:result_compnocom} (a) we compared the FDP solution with ZF \cite{8417645} and \gls{CB} \cite{8417645}, and in Fig.~\ref{fig:result_compnocom} (b), the proposed \gls{HBF} solution has been compared with PE-AltMin as near-optimal solution. We consider different noise power values $\sigma^2$ ranging from $-110$~dBW to $-130$~dBW. When considering the channel attenuation, the average \glspl{SNR} are between $3.1$~dB and $23.1$~dB.
It can be seen that the proposed HBF and FDP architectures all provide near-optimal sum-rate performance with decentralized architecture over this noise power range. \hamed{Among FDP solutions}, the \gls{CB} has poor performance in a high SNR regime because user interference is more dominant. On the other hand, \gls{ZF} has poor performance in the low SNR regime and a low number of \glspl{AP}. Finally, Fig.~\ref{fig:result_CF_CDF} shows the cumulative distribution function~(CDF) of the per-user rates. 
It is shown that both proposed DNN architectures focus on maximizing the average sum-rate and neglect users with worse channels. This is expected since no notion of fairness has been included in the loss function used to train the DNNs.

\section{Conclusion} \label{sec:Conclusion}
\gls{CF-mMIMO} is a promising technique to increase the throughput and improve the coverage, but \francois{conventional approaches for designing the precoder at each AP are complex and require a significant communication overhead, both in the case of FDP and HBF architectures.}
In this paper, we proposed two RSSI-based \glspl{DNN} with distributed architectures \francois{to design a coordinated FDP or HBF precoder}. 
\hamed{The experimental results} \francois{on a millimeter-wave ray-tracing model} \hamed{show that the proposed DNNs can achieve near optimal performance for both FDP and HBF systems, while significantly reducing the computational complexity and the signaling overhead. Furthermore, the signaling overhead can be completely eliminated at the cost of increased complexity.}

\bibliographystyle{IEEEtran}
%
\bibliography{bib/HBF_bib.bib}

\end{document}